\documentclass{article}
\usepackage{graphics,latexsym}
\usepackage{graphicx}
\usepackage{amsmath, amssymb,amsthm}

\begin{document}

\title{Numerical Simulation \\ of Electromagnetic Solitons
\\ and their Interaction with Matter}
\date{}
\maketitle

\centerline{Daniele Funaro}
\centerline{Department of Mathematics, University of Modena and Reggio Emilia}
\centerline{Via Campi 213/B, 41100 Modena (Italy)}
\centerline{daniele.funaro@unimore.it}

\begin{abstract}
A suitable correction of the Maxwell model  brings to an
enlargement of the space of solutions, allowing for the existence
of solitons in vacuum. We review the basic achievements of the
theory and discuss some approximation results based on an explicit
finite-difference technique. The experiments in two dimensions
simulate travelling solitary electromagnetic waves, and show their
interaction with conductive walls. In particular, the classical
dispersion, exhibited by the passage of a photon through a small
aperture, is examined.
\end{abstract}



\renewcommand{\theequation}{\thesection.\arabic{equation}}

\section{Introduction}
\label{sec1} It is known, although rarely stated in books, that
Maxwell equations in empty space do not admit finite-energy
solitary waves among their solutions. By this we mean smooth
electromagnetic waves with compact support, travelling along
straight-lines at the speed of light. One of the reasons to
explain this fact can be attributed to the linearity of the
Maxwell model, not allowing the due focussing of the signal on a
constrained path. Such an impediment has stimulated the research
of alternative nonlinear models, mostly based on modification of
the Lagrangian (see for instance \cite{borni}),  granting the
existence of soliton-like solutions and justifying in this way
electromagnetic phenomena such as photons.
\par\smallskip

A step ahead in the comprehension of electromagnetic  solitary
waves has been done in \cite{funaro} and \cite{funarol}, where, in
the framework of a self-consistent theory, explicit analytic
solutions are carried out. The main argument is that the classical
equations of the electromagnetism in vacuum are not capable to
follow the evolution of finite-energy wave-fronts in the proper
way (i.e., the one described for instance by the Huygens
principle); this is due to the difficulty to impose both ${\rm
div}{\bf E} =0$ and ${\rm div}{\bf B} =0$ at the same time on each
point of the same front. As a matter of fact, if  we define the
wave-front as the surface envelope of the vector fields, one can
easily show that requiring that the front evolves along its normal
direction is incompatible with the two free-divergence conditions
taken together. Therefore, the relation ${\rm div}{\bf E} =0$ has
been dropped, without this necessarily implying the existence of
point-wise electric charges. Here, we circumscribe our exposition
by outlining some of the peculiarity of the new approach, while we
refer to \cite{funarol} for a detailed explanation of the origin
and the possible physical implications.
\par\smallskip

In particular, in this paper we devote our attention to some
numerical simulations. It is soon evident that the removal of
relation ${\rm div}{\bf E}=0$  is important for the  construction
of numerical algorithms in general, because most of the
difficulties encountered in simulations are  indeed consequence of
the imposition of such a constraint. These include the efforts
made to build approximation spaces satisfying some divergence-free
conditions (see for instance \cite{assous}, \cite{boffi},
\cite{bossavit}, \cite{cockburn}, \cite{hyman}, \cite{monk}) or
divergence corrections techniques (see for instance \cite{konrad},
\cite{munz}, \cite{rahman}). Some modifications of the Maxwell
model, which may be in some way assimilated to the ones considered
here, have been proposed at numerical level, in order to set up
stable schemes (see  \cite{taflove}, \cite{yee}), to handle
boundary conditions (see  \cite{abarbanel}, \cite{berenger}), or
for the treatment of wave propagation in linear non-dispersive
lossy materials. These techniques are mainly adopted to overcome
numerical troubles and are not intended to modifying the Maxwell
model itself, as we are doing here. A survey of methods is given
in \cite{schilders}, where the reader can also find an updated
list of references.
\par\smallskip

For simplicity, we will use explicit finite-differences (in
particular the Lax-Wendroff method). The aim is to validate the
theory in \cite{funarol} with a series of simple experiments. To
this purpose we follow the evolution of solitary waves and see how
they react when encountering matter. We will mainly concentrate
to the physical appearance and significance  of the solutions.
Thus, no theoretical issues will be discussed. Moreover, the
discretization parameters will be small enough to get sufficiently
accurate approximations, finalized to recover qualitative
information. In no way we shall claim that our numerical approach
is competitive, being conscious that the one proposed here can be
certainly improved upon, as far as performances versus costs are
concerned.
\par\smallskip

The author would like to thank A. Ugolini (see \cite{ugolini})
for setting up the computational code.

\par\smallskip
\setcounter{equation}{0}
\section{The model equations}

\par\smallskip
By denoting with $c$  the speed of light, the classical Maxwell
equations, in void three-dimensional space, in absence of electric
charges, take the form:
\begin{equation}\label{eq:rotb}
 \frac{\partial {\bf E}}{ \partial t}~=~ c^2 {\rm curl} {\bf B}
\end{equation}
\begin{equation}\label{eq:dive}
{\rm div}{\bf E} ~=~0
\end{equation}
\begin{equation}\label{eq:rote}
\frac{\partial {\bf B}}{\partial t}~=~ -{\rm curl} {\bf E}
\end{equation}
\begin{equation}\label{eq:divb}
{\rm div}{\bf B} ~=~0
\end{equation}
where the two fields ${\bf E}$ and $c{\bf B}$ have the same
dimensions.
\par\smallskip

In \cite{funarol}, part of the analysis is devoted to
electromagnetic {\it free-waves}. These are solutions of the
following set of model equations:
\begin{equation}\label{eq:sfem2}
\frac{\partial {\bf E}}{\partial t}~=~ c^2{\rm curl} {\bf B}~
-~\rho {\bf V}
\end{equation}
\begin{equation}\label{eq:sfbm2}
\frac{\partial {\bf B}}{\partial t}~=~ -{\rm curl} {\bf E}
\end{equation}
\begin{equation}\label{eq:sfdb2}
{\rm div}{\bf B} ~=~0
\end{equation}
\begin{equation}\label{eq:slor2}
{\bf E}~+~{\bf V}\times {\bf B} ~=~0
\end{equation}
where $\rho ={\rm div}{\bf E}$, and ${\bf V}$ is a velocity
vector field satisfying $\vert{\bf V}\vert =c$. The field ${\bf
V}$ is oriented as the vector field ${\bf E}\times {\bf B}$. Note
that relation (\ref{eq:slor2}) is certainly satisfied for all
electromagnetic waves where ${\bf E}$ is orthogonal to ${\bf B}$
and $\vert {\bf E}\vert =\vert c{\bf B}\vert$. These requirements
are standard. In addition, we expect the new set of equations to
admit a large space of solutions. We will check later that,
contrary to the Maxwell model, solitary waves with compact
support, as well as perfect spherical waves, are now among the
solutions. On the other hand, if $~{\rm div}{\bf E}~$ is
relatively small (as it actually happens in many practical
circumstances), then (\ref{eq:sfem2}) is as accurate as the
corresponding standard Maxwell equation. Therefore, we expect the
new model to be consistent with the existing ones, for a broad
range of applications.
\par\smallskip

It should be clear that the added term $\rho {\bf V}=c\rho {\bf
J}$, where $\vert {\bf J}\vert =1$, has the meaning of a  current
density flowing in the direction of the rays at speed $c$
(Amp\`ere term). We assume this to be true even if there is no
presence of point-wise electric  charges. As documented in
\cite{funarol}, such a current is part of the wave itself.
Moreover, equation (\ref{eq:slor2}) actually characterizes
free-waves, since its says that there are no external `forces'
acting on the wave-fronts, as a result of external perturbations.
It is a sort of Lorentz law (where the force turns out to be
zero), in which moving charges are replaced by a balance of pure
vector fields. We will see later, in section 4, how  equation
(\ref{eq:slor2}) should be modified in order to handle situations
in which the wave interacts with matter.
\par\smallskip

It is important to observe that a conservation principle
holds. In fact, by taking the divergence of equation
(\ref{eq:sfem2}), we come out with the following continuity
equation:
\begin{equation}\label{eq:conti}
\frac{\partial \rho}{\partial t}~=~-c~{\rm div}(\rho {\bf J})
\end{equation}
In addition, using that ${\bf V}$ is orthogonal to both ${\bf E}$
and ${\bf B}$, one can easily prove the classical Poynting
relation:
\begin{equation}\label{eq:poyn}
\frac{1}{2} \frac{\partial }{\partial t}(\vert {\bf E}\vert^2 +\vert c{\bf B}\vert^2)
~=~-c^2~{\rm div}({\bf E}\times{\bf B})
\end{equation}

Such  properties confirm that what we are doing has physical
meaning. It can be proven that the electromagnetic free-waves
described by the new set of equations are perfectly compatible
with the Huygens principle and the eikonal equation (see for
instance \cite{born}). In fact, if, for example, ${\bf V}$ is
irrotational (i.e.: the rays are straight-lines), then ${\bf
V}=\nabla\Psi$ for some potential $\Psi$, so that the relation
$\vert {\bf V}\vert =c$ is equivalent to $\vert \nabla\Psi\vert
=c$, which is the eikonal equation. Moreover, (\ref{eq:sfem2})
follows from minimizing the standard Lagrangian of classical
electromagnetism, after imposing the constraint ${\bf A}=\Phi{\bf
J}$ to the electromagnetic potentials ${\bf A}$ and $\Phi$ (see
\cite{funaro2} and \cite{funarol}, section 2.4). The covariant
version of the equations and their invariance under Lorentz
transformations are also discussed in \cite{funarol}, chapter 4.

\par\smallskip
\setcounter{equation}{0}
\section{The discretization method for 2-D solitons}

Explicit expressions of wave-packets, moving at  speed $c$ and
solving the system of equations (\ref{eq:sfem2})-(\ref{eq:slor2}),
are available.  The electromagnetic signals carried by these
fronts can be very general. For the sake of simplicity, only
examples in two dimensions will be examined here. This does not
obscure, however, the general validity of the formulation.
\par\smallskip

We start by discussing a simple case. In the 3-D space with
cartesian coordinates $(x,y,z)$, we can define the following
fields:
$$
{\bf E}~=~\Big( cf(x)g(ct-z),~0,~0\Big)
$$
\begin{equation}\label{eq:campi}
{\bf B}~=~\Big( 0,~ f(x)g(ct-z),~0\Big)~~~~~~~~{\bf V}~=~\big( 0,~0,~ c \big)
\end{equation}

Now, the continuity equation (\ref{eq:conti}) becomes the simple
transport equation:
\begin{equation}\label{eq:contis}
\frac{\partial \rho}{\partial t}~+~c\frac{\partial \rho}{\partial
z}~=~0
\end{equation}

In practice, the example is two-dimensional since there is no
dependency on the variable $y$. The fields describe the evolution
of a solitary wave-packet shifting along the axis $z$. This is a
group of parallel transversal fronts, travelling in the same
direction ${\bf V}$ and modulated in intensity by the function
$g$. It is important to remark that the fields ${\bf E}$ and ${\bf
B}$ satisfy Maxwell equations (including (\ref{eq:dive})) only if
the function $f$ is constant (this check can be done by directly
differentiating the expression in (\ref{eq:campi})  and
substituting in the model equations). This corresponds to a plane
wave of infinite energy, which is the only Maxwellian front
allowed by the above setting. If we want the energy to be finite,
the only possible choice is $f=0$, implying ${\bf E}=0$. The
situation is different if we take into account equation
(\ref{eq:sfem2}). Now, as the reader can easily check by direct
computation, the fields in (\ref{eq:campi}) are solutions, and the
two functions $f$ and $g$ can be completely arbitrary. In
particular,  $f$ and $g$ may have compact support. Note that, for
simplicity, the wave is unbounded in the $y$ direction, hence, in
this case the energy of the wave-packet is not finite. Examples
where the fronts are also bounded in the $y$ direction can be
easily constructed (recall to enforce the condition ${\rm div}{\bf
B} =0$), but the problem becomes three-dimensional and here we
would prefer to avoid this complication.
\par\smallskip

Let us discuss a specific example. We start by fixing a point
$(x_0,z_0)$. At time $t=0$, we can assign the distribution:
$$
{\bf E}~=~\Big( \phi_\tau(x)\phi_\sigma(z),~ 0,~0\Big)~~~~~~~~
{\rm with}
$$
\begin{equation}\label{eq:campoei}
\phi_\tau (x)~=~1+\cos \left(\frac{x-x_0}{\tau}\pi\right),~~~~~~~
\phi_\sigma (z)~=~1+\cos \left(\frac{z-z_0}{\sigma}\pi\right)
\end{equation}
inside the rectangle centered at $(x_0,z_0)$ with sides of length
$2\tau$ and $2\sigma$. Outside the same rectangle ${\bf E}$ is
required to vanish. Accordingly, field $c{\bf B}$ is supposed to
be orthogonal to the plane $(x,z)$ and with the same intensity as
${\bf E}$. When time passes, this signal-packet moves at speed $c$
in the direction of the axis $z$:
\begin{equation}\label{eq:campoev}
{\bf E}~=~\Big( \phi_\tau(x)\phi_\sigma(ct-z),~ 0,~0\Big)
\end{equation}

We do not need to use any approximation  technique in order to
follow the evolution of such a soliton, since the corresponding
analytic expression exactly solves the set of equations
(\ref{eq:sfem2})-(\ref{eq:slor2}). However, in view of more
complex applications, let us introduce the Lax-Wendroff method.
More precisely, we work with its 2-D version applied to a system
of pure hyperbolic equations, as described in \cite{strikwerda},
section 4.9.
\par\smallskip

Let us suppose that, for some real constant $d\times d$ matrices
$M_x$ and $M_z$, our system can be written in the following form:
\begin{equation}\label{eq:eqipe}
\frac{\partial {\bf u}}{\partial t}~=~M_x\frac{\partial {\bf u}}{\partial x}
~+~M_z\frac{\partial {\bf u}}{\partial z}
\end{equation}
where the $d$-components vector ${\bf u}$ depends on $x$, $z$ and $t$.
\par\smallskip

We denote by $(x_i,z_j)$ the nodes of a uniform grid (of step-size
$h$) on the plane $(x,z)$. The time-discretization parameter is
denoted by $\Delta t$. Based on the stencil shown in figure 1, we
pass from the approximated solution ${\bf u}^k$ at time $t_k$ to
the one at time $t_{k+1}=t_k+\Delta t$.

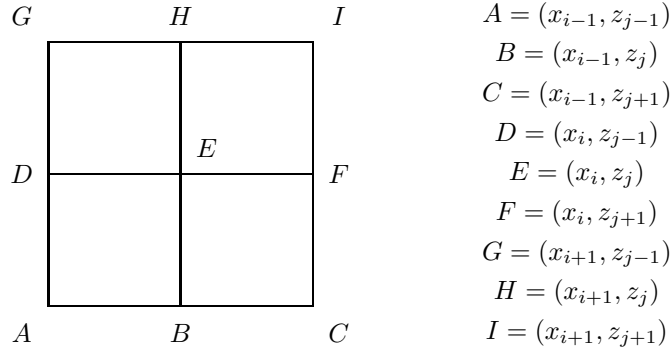
\begin{figure}[h]\hspace{.8cm}
\begin{picture}(300,180)
\put(50,50){\line(0,1){100}}
\put(50,50){\line(1,0){100}}
\put(50,100){\line(1,0){100}}
\put(50,150){\line(1,0){100}}
\put(100,50){\line(0,1){100}}
\put(150,50){\line(0,1){100}}
\put(40,40){\makebox(0,0){$A$}}
\put(160,40){\makebox(0,0){$C$}}
\put(100,40){\makebox(0,0){$B$}}
\put(40,160){\makebox(0,0){$G$}}
\put(160,160){\makebox(0,0){$I$}}
\put(100,160){\makebox(0,0){$H$}}
\put(40,100){\makebox(0,0){$D$}}
\put(160,100){\makebox(0,0){$F$}}
\put(110,110){\makebox(0,0){$E$}}
\put(250,160){\makebox(0,0){$A=(x_{i-1},z_{j-1})$}}
\put(250,145){\makebox(0,0){$B=(x_{i-1},z_{j})$}}
\put(250,130){\makebox(0,0){$C=(x_{i-1},z_{j+1})$}}
\put(250,115){\makebox(0,0){$D=(x_{i},z_{j-1})$}}
\put(250,100){\makebox(0,0){$E=(x_{i},z_{j})$}}
\put(250,85){\makebox(0,0){$F=(x_{i},z_{j+1})$}}
\put(250,70){\makebox(0,0){$G=(x_{i+1},z_{j-1})$}}
\put(250,55){\makebox(0,0){$H=(x_{i+1},z_{j})$}}
\put(250,40){\makebox(0,0){$I=(x_{i+1},z_{j+1})$}}
\end{picture}\vspace{-1.cm}
\caption{\small Stencil for the Lax-Wendroff method.}
\end{figure}

\medskip
\noindent This is done according to the scheme:
$$
{\bf u}^{k+1}(E)~=~\Big[ I-\lambda^2 (M_x^2+M_z^2)\Big]{\bf u}^k(E)~+~
{\textstyle{\frac{1}{2}}}\lambda M_x(I+\lambda M_x){\bf u}^k(H)
$$
$$
~-~{\textstyle{\frac{1}{2}}}\lambda M_x(I-\lambda M_x){\bf u}^k(B)
~+~{\textstyle{\frac{1}{2}}}\lambda M_z(I+\lambda M_z){\bf u}^k(F)
~-~{\textstyle{\frac{1}{2}}}\lambda M_z(I-\lambda M_z){\bf u}^k(D)
$$
\begin{equation}\label{eq:eqiped}
~+~{\textstyle{\frac{1}{8}}}\lambda^2 (M_xM_z+M_z M_x)({\bf u}^k(I)
-{\bf u}^k(G)-{\bf u}^k(C)+{\bf u}^k(A))
\end{equation}
where $~\lambda =\Delta t/h~$ and $I$ is the identity matrix.
\par\smallskip
Let us examine the case of a wave with the electric field laying
on the plane $(x,z)$. Therefore, the magnetic field, which is
orthogonal to that plane, has only one component different from
zero. The condition (\ref{eq:divb}) is always satisfied, since our
functions do not depend on the variable $y$. We organize the
vector ${\bf u}$ in the following way ($d=3$):
$${\bf u}~=~(E_1, ~cB_2, ~E_3)$$
For the classical Maxwell system (see (\ref{eq:rotb}) and (\ref{eq:rote}))
it is enough to define the matrices:
\begin{equation}\label{eq:matricim}
M_x~=~\left(\begin{array}{ccc} 0 & 0 & 0 \\ 0 & 0 & c \\0 & c & 0 \\\end{array}
\right)~~~~~~~~
M_z~=~\left(\begin{array}{ccc} 0 & -c & 0 \\ -c & 0 & 0 \\0 & 0 & 0 \\\end{array}
\right)
\end{equation}
in order to recover (\ref{eq:eqipe}). It is important  to remark,
however, that the condition (\ref{eq:dive}) is not enforced (this
should be done at the initial time, although, as we mentioned,
fronts having ${\rm div}{\bf E}=0$ and ${\rm div}{\bf B}=0$ are
extremely rare and out of interest for our applications). Note
that the eigenvalues of both matrices are $\pm c$, confirming that
the system is of hyperbolic type.
\par\smallskip
Regarding the modified system (see (\ref{eq:sfem2}) and (\ref{eq:sfbm2})),
the above  matrices have to be replaced by:
\begin{equation}\label{eq:matrici}
M_x~=~\left(\begin{array}{ccc} -V_1 & 0 & 0 \\ 0 & 0 & c \\-V_3 & c & 0 \\\end{array}
\right)~~~~~~~~
M_z~=~\left(\begin{array}{ccc} 0 & -c & -V_1 \\ -c & 0 & 0 \\0 & 0 & -V_3 \\\end{array}
\right)
\end{equation}
where ${\bf V}=(V_1,0,V_3)$ is a velocity vector orthogonal to
${\bf E}=(E_1,0,E_3)$, such that $~V_1^2+V_3^2=c^2$. Moreover, we
require relation (\ref{eq:slor2}) to be verified. In other terms,
${\bf V}$ must be defined as: $~c({\bf E}\times{\bf B})/\vert {\bf
E}\times{\bf B}\vert$. Note that this time we do not need to
satisfy (\ref{eq:dive}). Finally, let us observe that, for both
the matrices in (\ref{eq:matrici}), the modulus of the maximum
eigenvalue turns out to be equal to $c$.
\par\smallskip

The matrices in (\ref{eq:matrici}) are not constant.
Nevertheless,we will apply the Lax-Wendroff method to the new
system also in this circumstance.  Therefore, when implementing
(\ref{eq:eqiped}) at the $k$-th step, the entries of the matrices
$M_x$ and $M_z$ will be evaluated at the central point of the
stencil of figure 1. We are aware of the fact that this correction
may deteriorate a bit the convergence properties of the method.
\par\smallskip

\noindent In order to guarantee stability, we must impose the
following constraint (CFL condition):
\begin{equation}\label{eq:consta}
\lambda ~\leq~ \frac{1}{2c\sqrt{2}}
\end{equation}
In our experiments we always used the maximum $\Delta t$ allowed by
(\ref{eq:consta}), i.e.: $\Delta t=h/2c\sqrt{2}$. With this choice
we never encountered problems concerning stability.
\par\smallskip

We start by testing the discretization method on the soliton
corresponding to the initial conditions in (\ref{eq:campoei}),
with $\sigma =\tau$.  The initial condition for the discrete
scheme is obtained by interpolation at the grid-points. It is
interesting to note that, with this setting, the second component
of ${\bf E}$ remains zero during the evolution. This is true both
for the exact solution (see (\ref{eq:campoev})) and the discrete
one. As a matter of fact, the scheme (\ref{eq:eqiped}) follows
from repeated applications of the three matrices $I$, $M_x$ and
$M_z$. In this situation we have $V_1=0$ and $V_3=c$. Every time
we apply the matrix $M_x$  to a vector of the form $(*,*,0)$, we
get a vector of the same form, since $E_1=cB_2$. The same is true
regarding the matrix $M_z$. It is also interesting to note that
relation (\ref{eq:slor2}) is preserved for any $k$.

\begin{center}
\begin{figure}[!h] \vspace{.6cm}
\centerline{\includegraphics[width=11.cm,height=6.2cm]{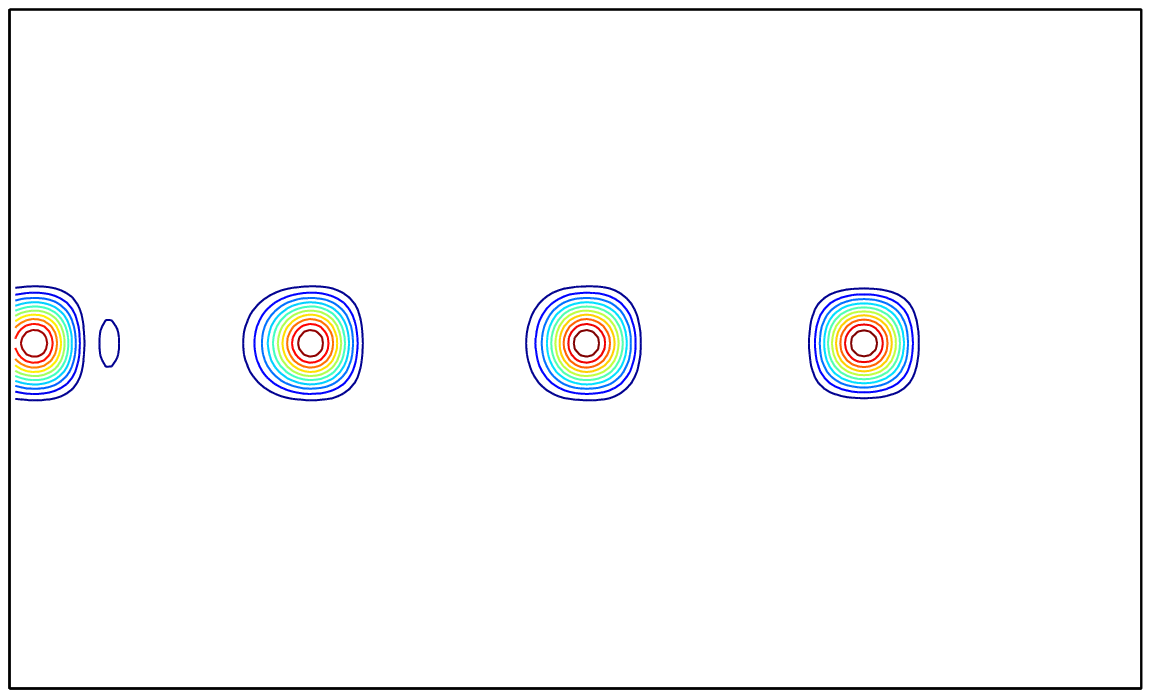}}\vspace{-.3cm}
\begin{caption}{\small Level sets of the scalar quantity $\vert{\bf E}\vert$, at different
times, for a free wave-packet shifting parallel to the $z$ axis
(from right to left). The vector ${\bf E}$ is orthogonal to the
direction of motion individuated by ${\bf V}$. The vector ${\bf
B}$ is orthogonal to the page.}
\end{caption}
\end{figure}
\end{center}
\vspace{-.6cm}

We can see in figure 2 the moving soliton at different times. The
discretization parameters have been taken small enough to get rid
of numerical disturbances. In other words, the parameter $\sigma$
is large enough to allow the inclusion of a  sufficiently high
number of grid-points in the support of the wave (there are
$13\times 13$ nodes in this example). It is interesting to remark
that the scalar function $E_1$ satisfies the wave equation:

\begin{equation}\label{eq:onde}
{\partial E_1\over \partial t^2}~=~c^2 {\partial E_1\over
\partial z^2}
\end{equation}

Note instead that the wave equation $~{\partial^2\over \partial
t^2}{\bf E}=c^2\Delta {\bf E}~$ does not hold, since in
(\ref{eq:onde}) we are omitting the derivatives with respect to
the variable $x$. On the other hand, we are not approximating a
trivial transport problem for the unknown $E_1$, but a coupled
hyperbolic system involving both ${\bf E}$ and ${\bf B}$.
This is not a trivial remark, but a rather important achievement.

\begin{center}
\begin{figure}[!h]\vspace{-.1cm}
\centerline{\includegraphics[width=5.cm,height=5.cm]{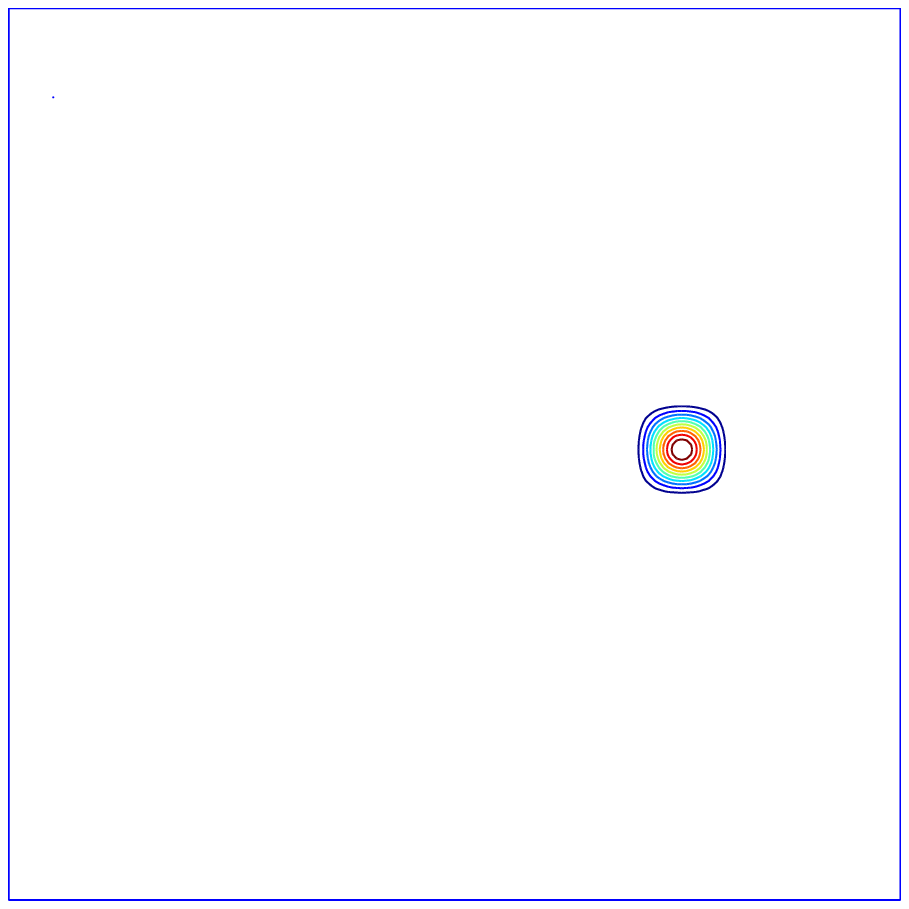}\hspace{1.3cm}
\includegraphics[width=5.cm,height=5.cm]{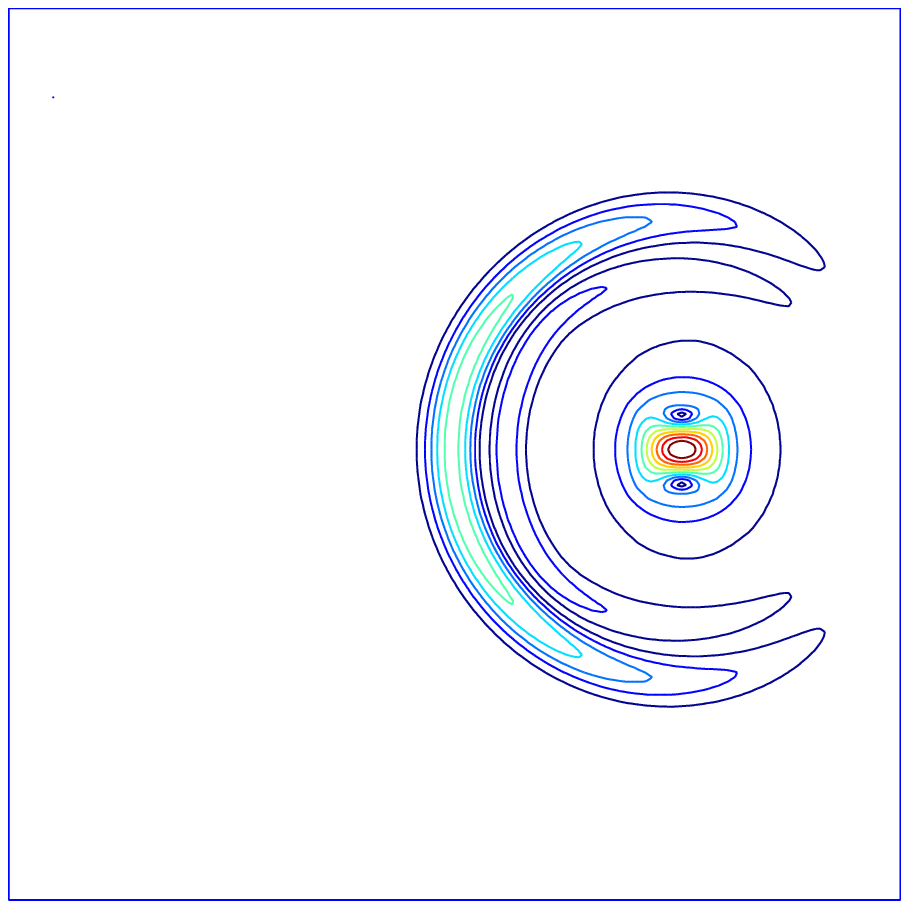}}\vspace{.6cm}
\centerline{\includegraphics[width=5.cm,height=5.cm]{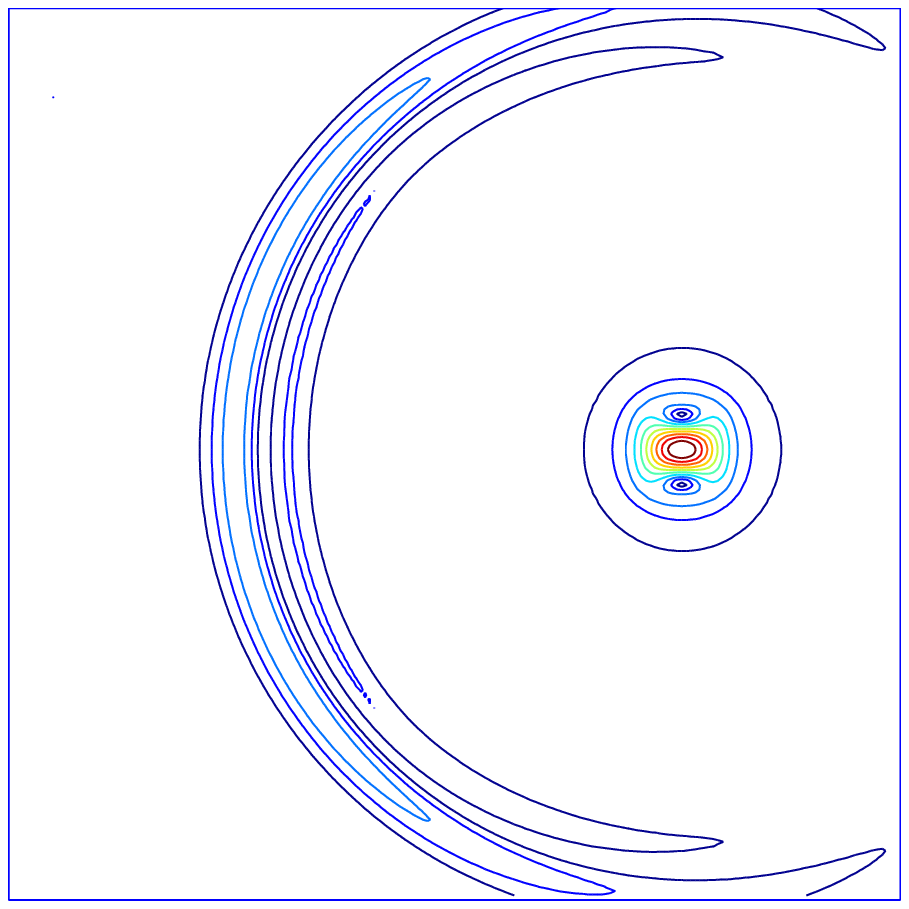}\hspace{1.3cm}
\includegraphics[width=5.cm,height=5.cm]{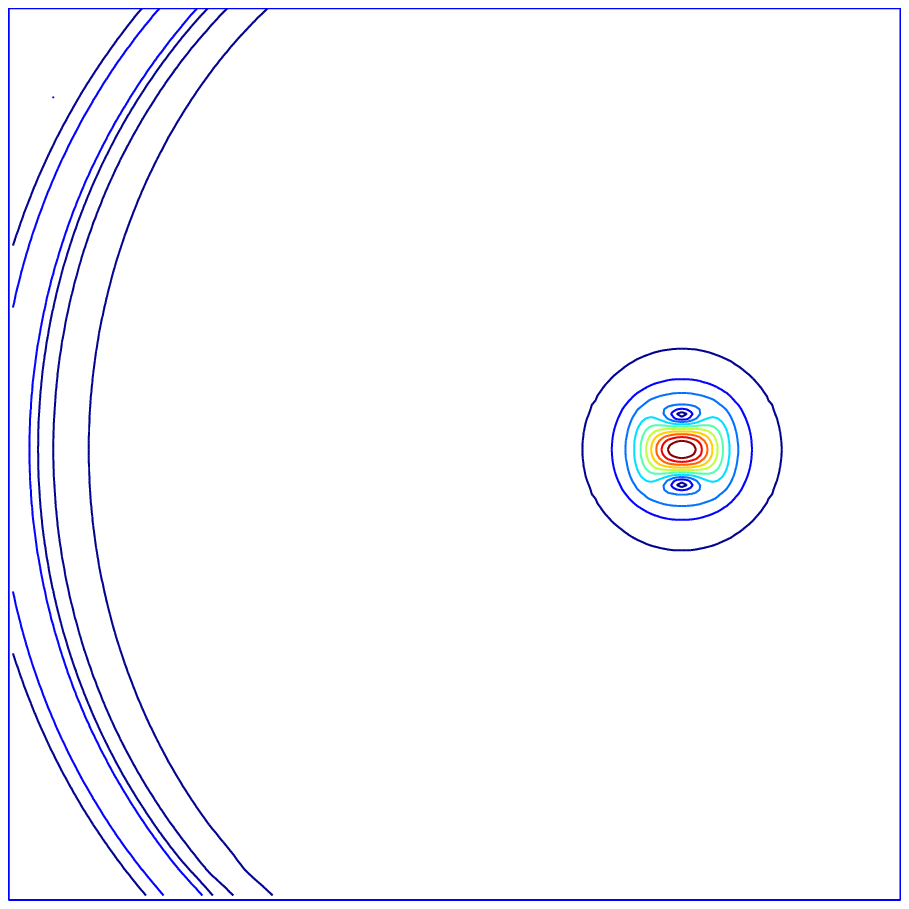}}
\begin{caption}{\small Level sets of $\vert{\bf E}\vert$, at different
times, obtained by implementing the scheme (\ref{eq:eqiped}) with
the matrices given in (\ref{eq:matricim}). }
\end{caption}
\end{figure}
\end{center}
\vspace{-.4cm}

Some longitudinal numerical dissipation is visible.  It can be
suitable reduced by using more appropriate numerical techniques.
It is important however to remark that no numerical viscosity is
introduced in the direction transverse to the motion, so that the
solitary wave proceeds maintaining a well defined width. This is
quite an important result. Suitable Sommerfeld type boundary
conditions are assumed at the outflow boundary, in order to allow
a smooth exit of the wave from the computational domain.
\par\smallskip

A completely different situation (see figure 3) is observed when
using the same initial datum (\ref{eq:campoei}) and evolving the
wave with the help of the two linear Maxwell equations
(\ref{eq:rotb}) and (\ref{eq:rote}), using the matrices in
(\ref{eq:matricim}). In this case the condition (\ref{eq:dive}) is
not verified, since such a constraint is not satisfied by the
initial datum and it is not enforced by the numerical scheme. As
the reader can see, part of the solution is diffused all around,
despite the fact that, in the initial dislocation, the vectors
${\bf V}$ are all oriented in the horizontal $z$ direction.
Surprisingly, the most of the energy does not move at all. Our
guess is that the condition ${\rm div}{\bf E}\not =0$ generates a
kind of stationary charge. Of course, this behavior is not
physically acceptable. Nevertheless, there is no way to modify the
initial datum, maintaining $f$ and $g$ with compact support. In
fact, as mentioned before, one can verify that in (\ref{eq:campi})
the only possible initial condition compatible with
(\ref{eq:dive}) is ${\bf E}=0$.  As far as solitons are concerned,
this indicates that the poor performances of the classical
equations of electromagnetism are mainly due to the total lack of
initial compatible conditions.

\begin{center}
\begin{figure}[h]\vspace{.3cm}
\centerline{\hspace{-.6cm}\includegraphics[width=3.cm,height=3.8cm]{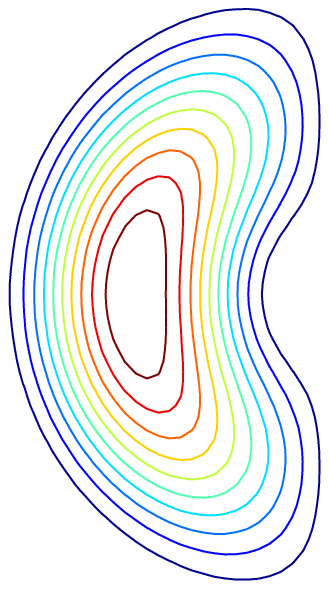}\hspace{.6cm}
\includegraphics[width=3.8cm,height=3.9cm]{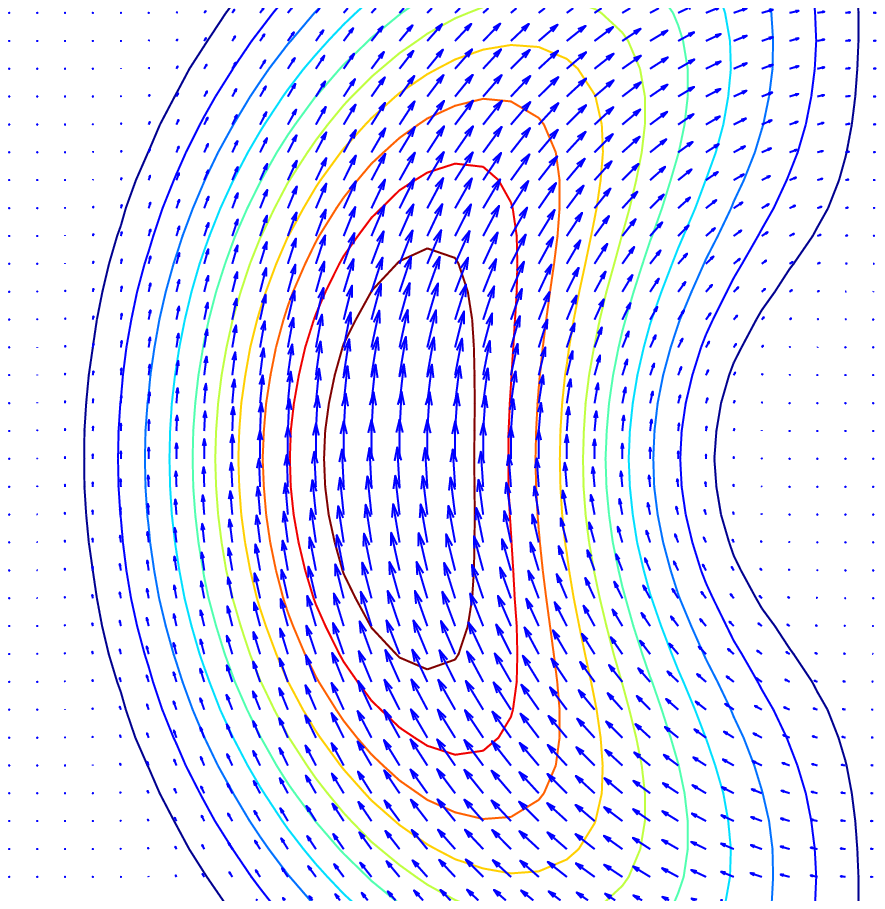}\hspace{.7cm}
\includegraphics[width=3.8cm,height=3.9cm]{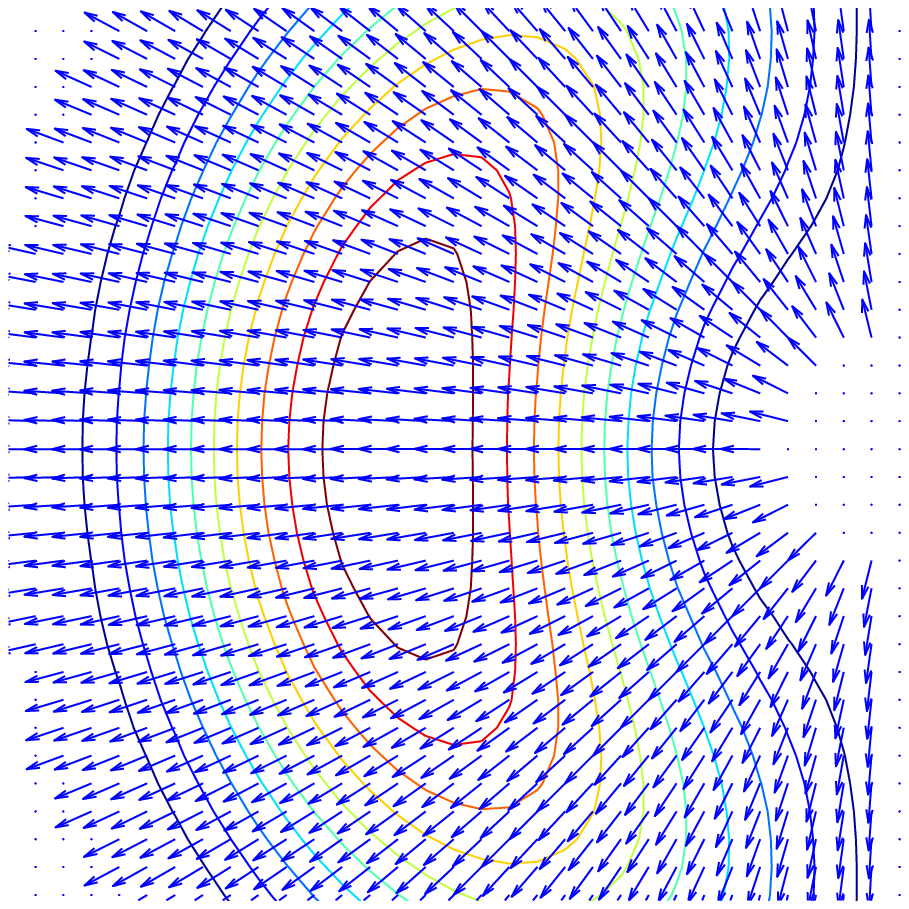}}
\begin{caption}{\small The first picture shows the level sets of $\vert
{\bf E}\vert$, for some initial datum of the form given by (\ref{eq:dato2}). The two
successive enlargements show the initial distributions of ${\bf
E}$ and ${\bf V}$ in the plane $(x,z)$.}
\end{caption}
\end{figure}
\end{center}\vspace{-.1cm}

A less trivial example is obtained by taking an initial datum of
the form:
\begin{equation}\label{eq:dato2}
{\bf E}~=~\Big(E_1(x,z), ~0, ~E_3(x,z)\Big)
\end{equation}
whose shape is given in figure 4. The magnetic field is along the
direction of the $y$-axis and such that $\vert {\bf E}\vert =\vert
c{\bf B}\vert$. The corresponding velocity field ${\bf V}$ is
radially distributed with respect to a prescribed point.

\begin{center}
\begin{figure}[h] \vspace{-.3cm}
\centerline{\includegraphics[width=13.cm,height=9.5cm]{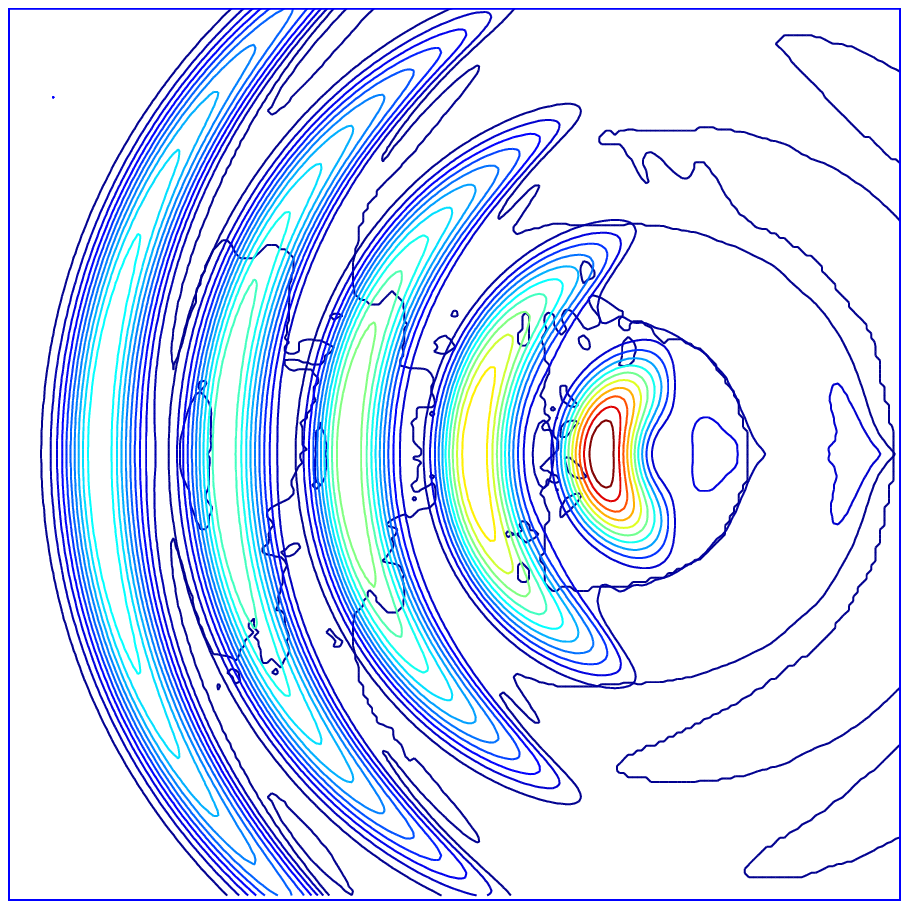}}\vspace{-.8cm}
\begin{caption}{\small Evolution of the wave corresponding to the
initial condition of figure 4, as obtained by running the
numerical scheme (\ref{eq:eqiped}) with a relatively small
discretization parameter.}
\end{caption}
\end{figure}
\end{center}

With this setting, during the evolution one gets a
semi-cylindrical wave,  simulated by the numerical scheme as shown
in figure 5. There is  a decay in intensity of the electromagnetic
fields, due to the fact that the signal is spread out on front
surfaces having increasing magnitude. This is in agreement with
the equation of continuity and the energy preservation rules (see
(\ref{eq:conti}) and (\ref{eq:poyn})), automatically imposed by
the governing equations. The approximation is not excellent, since
it is polluted by a secondary wave of small amplitude evolving
backwards. We think that this is due to the poor performances of
the Lax-Wendroff scheme. In fact, in the non-convex part of the
support of the wave some oscillations are generated. They modify
the sign of the electric field, but not the one of the magnetic
field, so that the associated velocity field ${\bf V}$
(proportional to the Poynting vector ${\bf E}\times{\bf B}$)
locally changes orientation. This anomalous part is then correctly
simulated by the numerical method, but, unfortunately it develops
along a wrong path. This is a minor effect that should be however
kept in mind when designing an alternative code.
\par\smallskip

After testing our algorithm on these simple examples, we are now ready to deal
with a more serious situation.

\par\smallskip
\setcounter{equation}{0}
\section{Constrained waves}

A generalization of the model is obtained by modifying equation
(\ref{eq:slor2}) in the  following way (see \cite{funarol},
chapter 5):
\begin{equation}\label{eq:sfgr2p}
\frac{D{\bf V}}{Dt}~=~-\mu\big({\bf E}~+~{\bf V}\times{\bf B}\big)
~-~\frac{\nabla p}{\rho}
\end{equation}
where the constant $\mu >0$ is a charge divided by a mass, $\rho
={\rm div}{\bf E}$ and $p$ is a kind of pressure. As customary, the substantial
derivative ${\bf G}=\frac{D}{Dt}{\bf V}$ is defined as
$\frac{D}{Dt}{\bf V}= \frac{\partial}{\partial t}{\bf V}+({\bf
V}\cdot \nabla ){\bf V}$, so that ${\bf G}$ turns out to be an
acceleration. Basically, if ${\bf V}$ is the normalized vector
field tangent to the light rays, then the field ${\bf G}$ gives a
measure of their curvature.
\par\smallskip

Equation (\ref{eq:sfgr2p}), even if there are no classical moving
charges, can be assimilated to the Lorentz law. If ${\bf G}$ and
$p$ are zero, then one is dealing with a free electromagnetic
wave, and the corresponding rays are straight-lines.  Actually,
equation (\ref{eq:slor2}), where ${\bf G}=0$ and $p=0$, says that
the evolution of the wave is free from constraints, that is, there
are no external factors (`forces') acting on it. When, for some
reasons, (\ref{eq:slor2}) is not satisfied, then pressure
develops. At the same time $\frac{D}{Dt}{\bf V}$ is different from
zero, so that ${\bf V}$ changes direction (the rays are curving)
and the electromagnetic wave-fronts locally follow the evolution
of the new normalized Poynting vector ${\bf J}={\bf V}/c$. Thus,
when ${\bf G}$ is different from zero, the wave is no longer free,
and, following \cite{funarol}, it will be called {\sl constrained
wave}. Such a situation  happens when the wave is subjected to
external electromagnetic fields. This is true for instance during
the interaction with matter at atomic level, as in reflection or
diffraction. Therefore, our model equations  are now able to
provide the coupling between the curvature of the rays and the
motion of the wave-fronts. Note that, during the change of
trajectory of the rays, depending on the context, the polarization
may also vary.
\par\smallskip

It is evident that equation (\ref{eq:sfgr2p}) is inspired by the
Euler equation for inviscid fluids (with an electromagnetic type
forcing term). Put in other words, the vector field ${\bf V}$
evolves according to the laws of fluid dynamics. At the same time,
it carries, on the transversal manifold, the two orthogonal fields
${\bf E}$ and ${\bf B}$. In this peculiar way, the system of
equations
(\ref{eq:sfem2})-(\ref{eq:sfbm2})-(\ref{eq:sfdb2})-(\ref{eq:sfgr2p})
describes the evolution of the triplet  $({\bf E}, {\bf B}, {\bf
V})$. We remark that our model is a system of vector equations.
Therefore, although our analysis is based on simple examples, it
has nothing to do with the equations modelling scalar solitons
(see for instance \cite{filippov}).
\par\smallskip

The exact physical meaning of  (\ref{eq:sfgr2p}) is carefully
explained in \cite{funarol}, where an equation of state involving
$p$, that uses the curvature of the space-time geometry, is also
introduced. For the sake of simplicity, we do not discuss further
the reasons of this construction and we refer to \cite{funarol}
for the details. Numerical experiments on constrained waves,
concerning photons trapped in toroid shaped regions, are
illustrated in \cite{chinosi}.

\par\smallskip
\setcounter{equation}{0}
\section{Some experiments on diffraction}

Our next step is to analyze the interaction of an incoming soliton
with a perfectly conducting plane wall. We would like to discuss
some experiments concerning diffraction, so that there is a hole
in the wall and our soliton will be forced to pass through it. The
barrier is simulated by imposing suitable boundary conditions, so
that the impact is immediate and discontinuities are generated.

\begin{center}\vspace{-.3cm}
\begin{figure}[!h]
\centerline{\includegraphics[width=9cm,height=8.2cm]{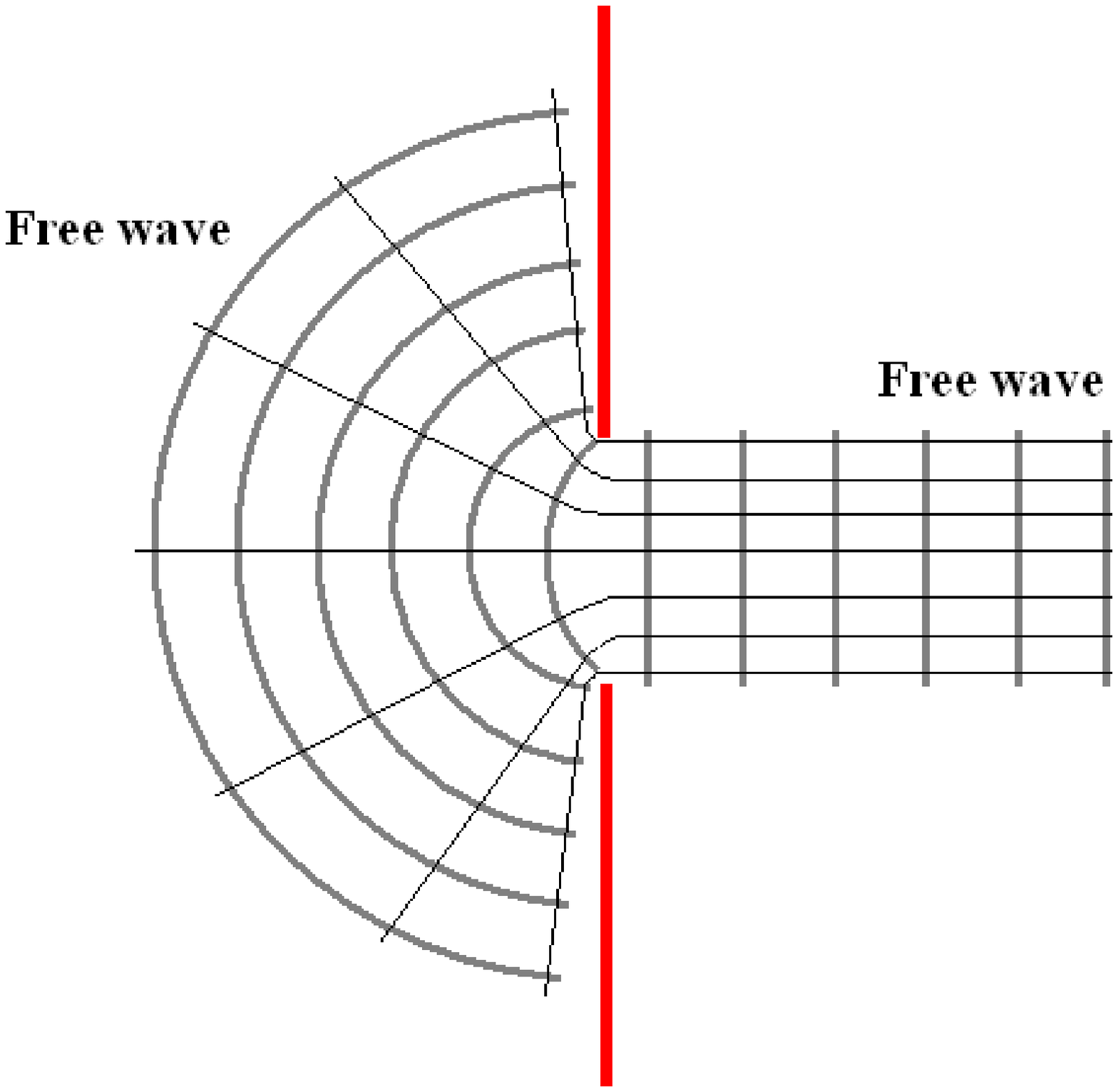}}\vspace{-.3cm}
\begin{caption}{\small An example of diffraction. A soliton is scattered
when passing through an aperture smaller than its size, producing
a diffusive effect.}
\end{caption}
\end{figure}
\end{center}

According to figure 6, we expect the soliton to behave as a
free-wave (hence ${\bf G}=0$ and $p=0$ in (\ref{eq:sfgr2p}))
before reaching the aperture. The reaction of the wall brings to
an instantaneous diffusion of the signal, as also observed in
real-life experiments with fluids or electromagnetic waves. After
the passage, the soliton continues his journey behaving as a free
wave, following however non-parallel characteristic
straight-lines.
\par\smallskip

Assuming that the path of the incoming wave is orthogonal to the
boundary, there we impose ${\bf V}=0$ (no slip condition). Hence,
for an instant, the last term on the right-hand side of equation
(\ref{eq:sfem2}) disappears. This way of imposing boundary
conditions is suggested in \cite{funarol}, chapter 5, where the
following Bernoulli type equation is deduced:
\begin{equation}\label{eq:berno}
\frac{D}{Dt}\left( p~+~ \frac{\rho}{2}\Vert {\bf V}\Vert^2\right)~=~-\frac{\rho}{2}
\Vert {\bf V}\Vert^2 ~{\rm div}{\bf V}
\end{equation}
Heuristically, when ${\bf V}$ goes to zero at  the boundary, the
pressure raises from zero to infinity, in order to maintain the
energy balance. One can actually check that $~{\rm div}{\bf V}>0~$
after the impact, bringing to a diffusive effect (with no
dissipation of energy, however). As we shall see from the
experiments, these assumptions seem to be correctly posed.

\begin{center}
\begin{figure}[!h]\vspace{-.4cm}
\centerline{\includegraphics[width=11.2cm,height=8.4cm]{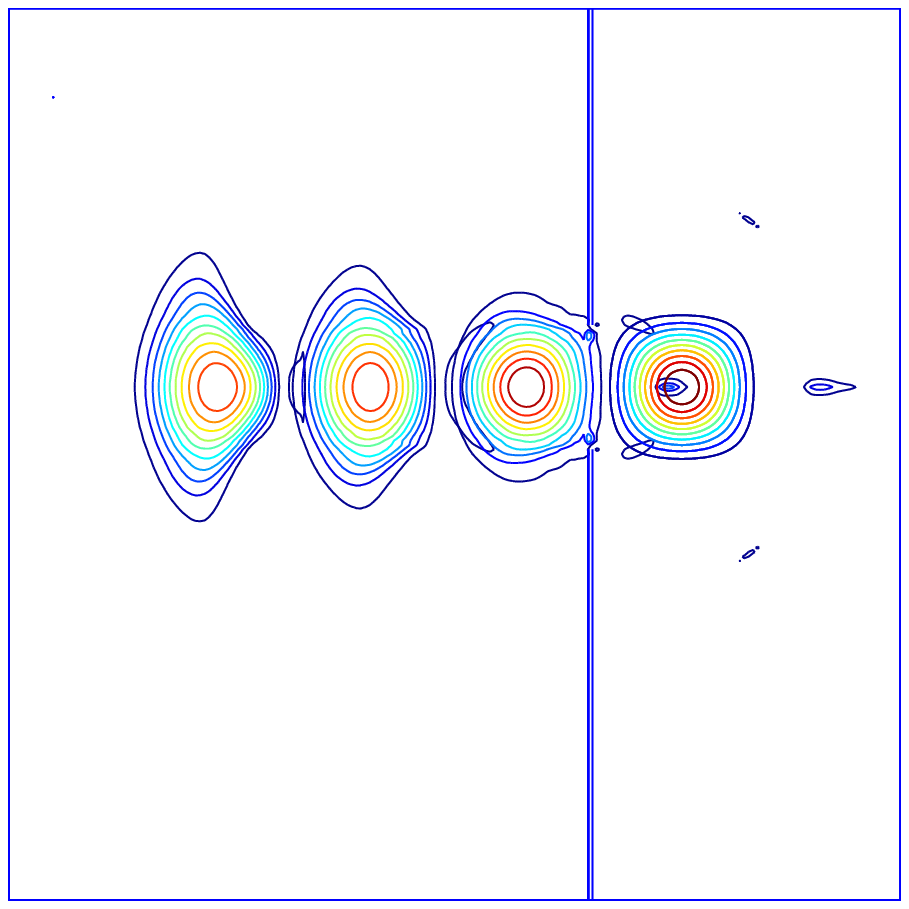}}\vspace{-.9cm}
\begin{caption}{\small Evolution of a solitary wave-packet passing through the
hole of a conductive wall slightly smaller than its width.}
\end{caption}
\end{figure}
\end{center}

From the numerical point of view the problem is {\it stiff},
therefore some unavoidable perturbations will be observed. A more
realistic behavior could be obtained by smoothing the effects of
the boundary. This could be simulated by slowing down the incoming
soliton by varying ${\bf V}$ in a continuous way. We did not
investigate however this possibility.

\begin{center}
\begin{figure}[!h]\vspace{-.4cm}
\centerline{\includegraphics[width=11.2cm,height=8.4cm]{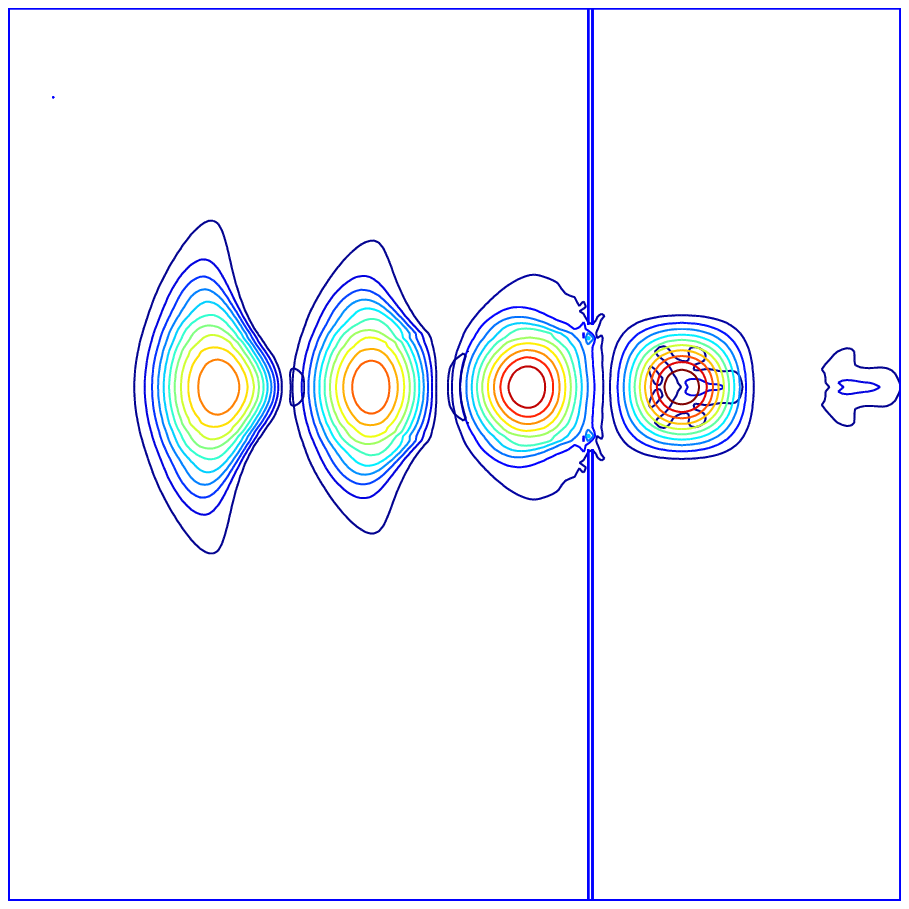}\vspace{-.6cm}}
\begin{caption}{\small Evolution of a solitary wave-packet passing through the
hole of a conductive wall, having smaller size if compared to the
one considered in figure 7.}
\end{caption}
\end{figure}
\end{center}

In figures 7 and 8, we can see the evolution of the norm of the
electric field. The width of the incoming soliton includes 32 grid
points, while the size of the hole corresponds to 28 grid points
in figure 7 and 26 grid points in figure 8. As expected, the no
slip condition at the boundary produces a diffusive behavior.
After hitting the obstacle, the development of the soliton is
similar to that of figure 5 (with much less emphasis, the fronts
assume a circular asset, and we know from the experiments
of section 3 that the scheme can support this situation). There is no
loss of energy or introduction of transversal dissipation. The
spreading is a consequence of the new geometrical distribution of
the rays, following non-parallel trajectories (in other words, as
in figure 6, the stream-lines associated to the velocity  vector
${\bf V}$ are diverging).
\par\smallskip

There are little wiggles scattered all around. These are mainly
due to the exceeding part of the wave, reflected back by the wall.
Therefore, these disturbances might be (in part) physically
consistent. Meanwhile, they can also point out the limit of the
algorithm and suggest better ways to handle boundary conditions
(especially for the portion involved in the reflection process).
\par\smallskip

Finally, in figure 9, we tested the case when the incoming soliton
is not symmetrically centered with respect to the aperture. Of
course, we could show a multitude of other scattering experiments,
regarding wave-packets of arbitrary shape interacting with a
variety of obstacles. However, the purpose here was to provide a
general qualitative insight.

\begin{center}
\begin{figure}[!h]\vspace{-.4cm}
\centerline{\includegraphics[width=11.2cm,height=8.4cm]{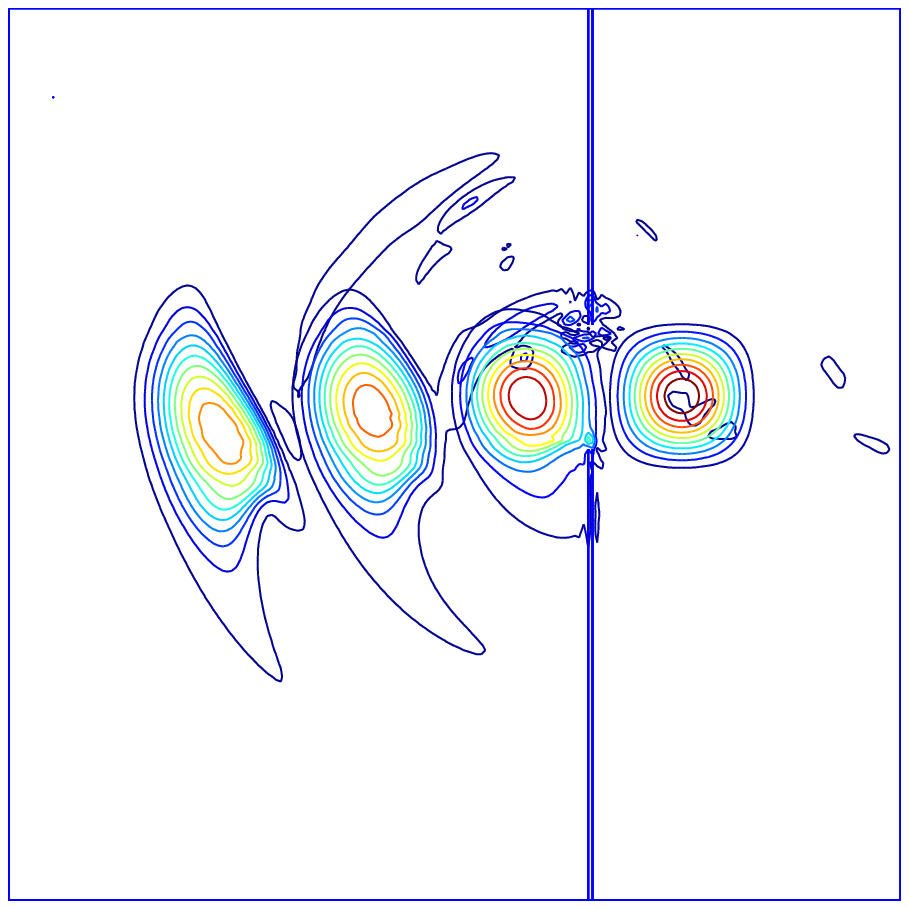}}\vspace{-.6cm}
\begin{caption}{\small Evolution of a solitary wave-packet passing through the
hole of a conductive wall. The aperture is  now asymmetrically
placed with respect to the trajectory of the wave-packet. Some
minor disturbances are present. To some extent, they can be
physically acceptable, but they are also due to the diffusive
effects of the Lax-Wendroff method, contrasted by the sharp
boundary constraints.}
\end{caption}
\end{figure}
\end{center}

These experiments can certainly be ameliorated by using more
appropriate approximations methods. Rather than accurate results,
the interest here was to show that, with the help of the model
equations
(\ref{eq:sfem2})-(\ref{eq:sfbm2})-(\ref{eq:sfdb2})-(\ref{eq:sfgr2p}),
an electromagnetic wave can be treated exactly as a material
fluid. We would like to remark once again that, although the
pictures presented here are related to the evolution of the scalar
quantity $\vert {\bf E}\vert$, we are not solving a trivial scalar
equation. Our solutions are electromagnetic localized emissions,
presenting both the features of waves and particles (photons).
Their internal structure is perfectly determined in terms of
electric and magnetic fields. In the more complex and realistic
3-D case, they can even change polarization, reacting in this way
like true electromagnetic entities,  but with energies
concentrated in finite regions of space. We believe that these
peculiarities are important both for physical implications and
technical applications.

\par


\smallskip

\begin{thebibliography}{99}

\bibitem{abarbanel} Abarbanel S., Gottlieb D., A mathematical analysis
of the PML method, J. Comput. Phys, {\bf 134}, p. 357-363, 1997.

\bibitem{assous} Assous F., Degond P., Heintze E., Raviart P. A., Segre J.,
On a finite-element method for solving the three dimensional
Maxwell equations, J. Comput. Phys., {\bf 109}, p. 222-237, 1993.


\bibitem{berenger} Berenger J.-P., A perfectly matched layer for the absorption
of electromagnetic waves, J. Comput. Phys., {\bf 114}, p. 185-200,
1994.

\bibitem{boffi} Boffi D., Fernandes P., Gastaldi L., Perugia I., Computational
models of electromagnetic resonators: analysis of edge element approximation,
SIAM J. Numer. Anal., {\bf 36}, p. 1264-1290, 1999.

\bibitem{borni}   Born M., Infeld L., Foundations of the new field theory,
Proc. R. Soc. Lond. A, {\bf 144}, p. 425-451, 1934.

\bibitem{born}   Born M., Wolf E., {\sl Principles of Optics},
Pergamon Press, Oxford, 1987.

\bibitem{bossavit} Bossavit A., {\sl Computational Electromagnetism}, Academic
Press, Boston, 1998.

\bibitem{chinosi} Chinosi C., Della Croce L.,
Funaro D., Rotating electromagnetic waves in toroid-shaped
regions,  to appear in Int. J. of Modern Phys. C.

\bibitem{cockburn}   Cockburn B., Shu C.-W., Locally divergence-free discontinuous
Galerkin methods for the Maxwell equations, SIAM J. Numer. Anal.,
{\bf 35}, p. 2440-2463, 1998.

\bibitem{filippov}   Filippov A. T., {\sl The Versatile Soliton}, Birkh\"auser, Boston, 2000.

\bibitem{funaro} Funaro D., A full review of the theory of
electromagnetism, arXiv/physics/0505068.

\bibitem{funarol} Funaro D., {\sl Electromagnetism and the Structure of Matter},
World Scientific, Singapore, 2008.

\bibitem{funaro2} Funaro D., A model for electromagnetic solitary waves in vacuum,
preprint.

\bibitem{hyman} Hyman J. M., Shashkov M., Natural discretization for the divergence,
gradient, and curl on logically rectangular grids, Computers Math.
Applic., {\bf 33}, n. 4, p. 81-104, 1997.

\bibitem{konrad}   Konrad A., A method for rendering 3D finite element vector
field solution non-divergent, IEEE Trans. Magnetics,
{\bf 25}, p. 2822-2824, 1989.

\bibitem{landau}   Landau L. D., Lifshitz E. M., {\sl The Classical Theory
of Fields}, Pergamon Press, Warsaw, 1962.

\bibitem{monk}   Monk P., {\sl Finite Elements Methods for Maxwell's
Equations}, Oxford Univ. Press, New York, 2003.

\bibitem{munz} Munz C.-D., Schneider R., Sonnendr\"ucker E., Voss U.,
Maxwell's equations when the charge conservation is not satisfied,
C. R. Acad. Sci. Paris, {\bf t.328}, S\'erie I, p. 431-436, 1999.

\bibitem{rahman}  Rahman B., Davies J., Penalty function improvement
of waveguide solution by finite elements, IEEE Trans. Microwave Theory and
Techniques, {\bf MTT-32}, p. 922-928, 1984.

\bibitem{schilders} Schilders W. H. A., ter Maten E. J. W. (Guest Editors),
{\sl Handbook of Numerical Analysis, Volume XIII, Numerical
Methods in Electromagnetics}, Ciarlet P. G. editor, Elsevier,
2005.

\bibitem{strikwerda} Strikwerda J. C., {\sl Finite Difference Schemes and Partial
Differential Equations}, Wadsworth and Brooks, CA, 1989.

\bibitem{taflove} Taflove A.,  Hagness S. C., {\sl Computational Electrodynamics:
The Finite-Difference Time-Domain Method}, Artech House, Norwood
MA, 2000.

\bibitem{ugolini} Ugolini A., Thesis: Approssimazione di Onde
Elettromagnetico--Gra\-vitazionali, Universit\`a di Modena e
Reggio Emilia, 2007.

\bibitem{yee} Yee K. S., Numerical solution of initial boundary value problems
involving Maxwell's equations in isotropic media, IEEE Trans.
Antennas Propagat., {\bf 14}, p. 302-307, 1966.
\end{thebibliography}
\end{document}